\documentclass[twocolumn,showpacs,preprintnumbers,amsmath,amssymb,superscriptaddress]{revtex4}

\usepackage{graphicx}
\usepackage{dcolumn}
\usepackage{bm}

\begin{document}

\title{First direct measurement of the total cross section of 
$^{12}\rm C(\rm \alpha,\gamma)^{16}\rm O$} 

\author{D.Sch\"urmann}
 \affiliation{Institut f\"ur Experimentalphysik III,
	Ruhr-Universit\"at Bochum, Bochum, Germany}
\author{A.Di Leva}
 \affiliation{Institut f\"ur Experimentalphysik III,
	Ruhr-Universit\"at Bochum, Bochum, Germany}
\author{L.Gialanella}
	\affiliation{Dipartimento di Scienze Fisiche,
	Universit\`a Federico II, Napoli and INFN, Napoli, Italy}
\author{D.Rogalla}
	\affiliation{Dipartimento di Scienze Ambientali, Seconda
	Universit\`a di Napoli, Caserta and INFN, Napoli, Italy}	
\author{F.Strieder}
 	\affiliation{Institut f\"ur Experimentalphysik III,
	Ruhr-Universit\"at Bochum, Bochum, Germany}	
\author{N.De Cesare}
	\affiliation{Dipartimento di Scienze della Vita, Seconda
	Universit\`a di Napoli, Caserta and INFN, Napoli, Italy}
\author{A.D'Onofrio}
	\affiliation{Dipartimento di Scienze Ambientali, Seconda
	Universit\`a di Napoli, Caserta and INFN, Napoli, Italy}
\author{G.Imbriani}
 	\affiliation{Dipartimento di Scienze Fisiche,
	Universit\`a Federico II, Napoli and INFN, Napoli, Italy}
\author{R.Kunz}
	\affiliation{Institut f\"ur Experimentalphysik III,
	Ruhr-Universit\"at Bochum, Bochum, Germany}
\author{C.Lubritto}
	\affiliation{Dipartimento di Scienze Ambientali, Seconda
	Universit\`a di Napoli, Caserta and INFN, Napoli, Italy}
\author{A.Ordine}
	\affiliation{Dipartimento di Scienze Fisiche,
	Universit\`a Federico II, Napoli and INFN, Napoli, Italy}
\author{V.Roca}
	\affiliation{Dipartimento di Scienze Fisiche,
	Universit\`a Federico II, Napoli and INFN, Napoli, Italy}
\author{C.Rolfs}
 	\affiliation{Institut f\"ur Experimentalphysik III,
	Ruhr-Universit\"at Bochum, Bochum, Germany}
\author{M.Romano}
	\affiliation{Dipartimento di Scienze Fisiche,
	Universit\`a Federico II, Napoli and INFN, Napoli, Italy}
\author{F.Sch\"{u}mann}
 	\affiliation{Institut f\"ur Experimentalphysik III,
	Ruhr-Universit\"at Bochum, Bochum, Germany}
\author{F.Terrasi}
	\affiliation{Dipartimento di Scienze Ambientali, Seconda
	Universit\`a di Napoli, Caserta and INFN, Napoli, Italy}
\author{H.-P.Trautvetter}
 	\affiliation{Institut f\"ur Experimentalphysik III,
	Ruhr-Universit\"at Bochum, Bochum, Germany}

\begin{abstract}

The total cross section of $^{12}\rm C(\rm \alpha,\gamma)^{16}\rm O$ 
was measured for the first time by a direct and ungated detection
of the $^{16}\rm O$ recoils. This measurement in inverse
kinematics using the recoil mass separator ERNA in combination
with a windowless He gas target allowed to collect data with
high precision in the energy range $\rm E=1.9$ to 4.9~MeV.
The data represent new information for
the determination of the astrophysical S(E) factor.

\end{abstract}


\maketitle

\section{Introduction}

The radiative capture reaction $^{12}\rm C(\rm \alpha,\gamma)^{16}\rm O$
($\rm Q = 7.16$~MeV) takes place during stellar core helium burning \cite{1}, where 
$^{12}$C is produced by the triple-alpha process. The capture cross section
$\sigma(\rm E_0)$ at the relevant Gamow energy, $\rm E_0\approx0.3$~MeV
for $\rm T\approx2\times10^8$~K, determines -- together with the convection 
mechanism at the edge of the stellar core \cite{2} -- the abundances of carbon
and oxygen at the end of helium burning. 
This, in turn, strongly influences the nucleosynthesis of elements up to the
iron region for massive stars \cite{2} and the 
composition of CO white dwarfs, whose progenitors are intermediate
and low mass stars \cite{3}.

A recent experiment confirmed that the reaction rate of the triple-alpha
process is known with a precision of about 
10\% for temperatures near 10$^8$ K \cite{4}. A similar
precision is needed for the rate of $^{12}$C($\alpha,\gamma$)$^{16}\rm O$ to provide 
an adequate input for stellar models \cite{5}. The remarkable experimental
efforts over the last decades \cite{6,7,8,9,10,11,12,13,14} focused on 
the observation of the capture $\gamma$-rays, including one experiment \cite{9}
that used the coincident detection of the $^{16}\rm O$ recoils. Due to the 
low cross section and various background sources depending on the exact
nature of the experiments, $\gamma$-ray data with useful 
but still inadequate precision were limited to center-of-mass
energies ${\rm E_{cm} = E} = 1.0$ to 3.2~MeV. 
At the low-energy range, the data were limited e.g. by cosmic-ray
background and at the high-energy range the data were limited by intense
background reactions such as $^{13}$C($\alpha$,n)$^{16}$O (for an
$\alpha$-beam experiment) and $^{12}$C($^{12}$C,n)$^{23}$Mg
(for a $^{12}$C-beam experiment). 

The cross section $\sigma(\rm E_0)$ is expected to be dominated
by p-wave (E1) and d-wave (E2) capture to the ($\rm J^\pi = 0^+$) $^{16}\rm O$ 
ground state. Two bound states, at 6.92 MeV ($\rm J^\pi = 2^+$)
and 7.12~MeV ($\rm J^\pi = 1^-$), which correspond to subthreshold 
resonances at ${\rm E_R} = -245$ and $-45$~keV, appear to provide
the bulk of the capture strength $\sigma(\rm E_0)$ through their finite widths 
that extend into the continuum. R-matrix analyses 
are performed, in order to model the
energy dependence of the cross section. In these analyses the contribution of each
amplitude to the total cross section  is expressed in terms of 
a small number of resonances and a direct capture contribution. The 
parameters of the model are determined by a fit to the
experimental data. The extrapolation to $\rm E_0= 0.3$~MeV is 
sensitive to the properties of the nearby levels,
but it is sensitive also to the properties of the high lying 
resonances, since their tails extend to low energy. The
effect of these resonances is usually included by a
single high energy "background" resonance -- one for each
amplitude; these resonances are needed to obtain a 
good fit to the data.

Analyses of the available capture data together with data from the
$\alpha-^{12}$C elastic scattering and the $\beta$-delayed $\alpha$-decay
of $^{16}$N still lead to large uncertainties in the extrapolation
to E$_0$ \cite{22}. This is partly caused by large errors, both statistical 
and systematic, affecting the low energy capture data, and partly
due to the weak experimental constrains of the background 
resonances. Clearly, new measurements at low energies are
needed. However, of equal importance are also new 
measurements at significantly higher energies, well
above the range of the recent experiments, which 
may improve the experimental characterization of the background
resonances and may thus reduce the uncertainty in the extrapolated 
astrophysical S factor S(E$_0$). There may exist even an
s-wave capture (monopole E0) and significant capture 
amplitudes to excited $^{16}\rm O$ states. Since the various
capture amplitudes have different energy dependences, a better knowledge of
the individual amplitudes requires new data at both low and high
energies to provide an improved basis for their
extrapolation to E$_0$.

\section{Equipment and setup}

\begin{figure}
  \center\includegraphics[angle=90,width=8cm]{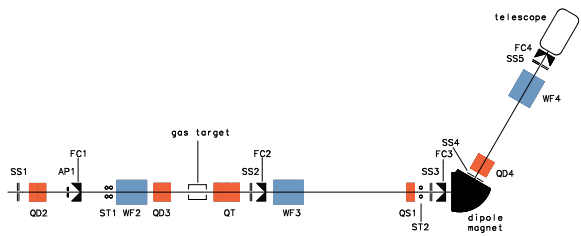}
  \vspace{-0.5cm}
  \caption{\label{setup}Schematic diagram of the recoil separator ERNA. WF =
  Wien filter, QS = quadrupole singlet, QD = quadrupole doublet, QT = quadrupole
  triplet, ST = steerer, FC = Faraday cup, SS = slit system, AP = aperture. At
  the end of the separator, there is a $\Delta$E-E telescope for particle
  identification.}
\end{figure}

A new experimental approach (fig. \ref{setup}) has been developed at the
4~MV Dynamitron tandem accelerator in Bochum, called 
ERNA = European Recoil separator for Nuclear Astrophysics \cite{15,16,17,18,19}.
In this approach, the reaction is initiated in 
inverse kinematics, $^4$He($^{12}$C,$\gamma$)$^{16}\rm O$, i.e. a $^{12}$C
ion beam with a particle current of up to 10~$\mu$A is guided into
a windowless $^4$He gas target. A beam contamination in the
order of $6\times10^{-10}$ $^{16}$O ions per incident $^{12}$C projectile
was found, by far too high for a direct detection of the $^{16}$O recoils.
Therefore, for the purpose of beam purification, there is one Wien filter 
before the analysing magnet (WF1, not shown in fig. \ref{setup})
and one before the gas target (WF2); with this combination a beam purity
better than 10$^{-18}$ could be achieved \cite{15}.
The windowless gas target - entrance and exit aperture
of the gas cell with a diameter of 6~mm -
includes an Ar post-target stripping system. After the
gas target, the separator consists sequentially of the 
following elements: a quadrupole triplet (QT), a Wien filter (WF3),
a quadrupole singlet (QS), a 60$^\circ$ dipole magnet, 
a quadrupole doublet (QD4), and a Wien filter (WF4).
The recoil separator suppresses
the intense $^{12}$C beam; the $^{16}\rm O$ recoils 
in a selected charge state are then counted in a $\Delta$E-E telescope
placed at the end of the beam line. The ratio between leaky
$^{12}$C events detected in the telescope and the number of $^{12}$C
projectiles is in the range of $4\times10^{-11}$ at the
high energy limit and $2\times10^{-13}$ at the lower limit.
Additionally, a beam suppression factor of the telescope alone of
10$^{-3}$ - ratio of $^{16}$O and leaky $^{12}$C in the $\Delta$E-E matrix
(fig. \ref{matrix}) - can be achieved leading to a total suppression factor of
better than $5\times10^{-14}$. ERNA 
is designed to study the reaction over the energy range
$\rm E = 0.7$ to 5.0~MeV. The detection of the $^{16}\rm O$ recoils allows, 
for the first time, a direct measurement of the total cross section of $^{12}\rm
C(\rm \alpha,\gamma)^{16}\rm O$, including possible non radiative 
transitions.

\section{Experimental procedures}

The number N$_{^{16}\rm O}$ of recoils collected in the telescope is given
by the relation

\begin{equation}
\rm N_{^{16}\rm O}= \rm N_{^4{\rm He}}\rm N_{^{12}\rm C}\sigma({\rm E_{eff}}){\rm \\
T_{RMS}}\Phi_{\rm R}\epsilon
\end{equation}
where N$_{^4{\rm He}}$ and N$_{^{12}\rm C}$ represent, respectively,
the target number density and the number of projectiles impinging on the 
target, $\sigma({\rm E_{eff}})$ is the cross section at the
effective interaction energy E$_{\rm eff}$, T$_{\rm RMS}$ is the transmission of ERNA
for the recoils in the selected charge state of
probability $\Phi_{\rm R}$, and $\epsilon$ is the detection 
efficiency of the telescope. Therefore $\sigma({\rm E_{eff}})$ can be
determined with high precision, if all quantities in 
Eq. 1 are known with high accuracy.

\begin{figure}
 \vspace{1cm}
  \center\includegraphics[angle=0,width=7cm]{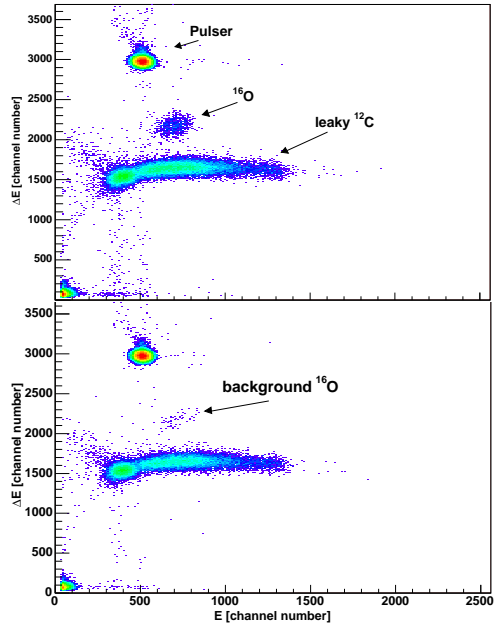}
  \caption{\label{matrix}Identification matrix of the $\Delta$E-E telescope at
  $\rm E=2.2$~MeV for the standard $^4$He target and Ar post-stripper density
  (upper panel). The different components are labeled. The lower panel shows
  the background at the same energy measured only with the Ar post-stripper
  after removing the $^4$He target gas. The running time for both spectra
  is the same.}
\end{figure}

The thickness of the extended windowless He gas target
was determined using the 
$^4$He($^7$Li,$\gamma$)$^{11}$B
reaction at ${\rm E_{lab}}=1.668$~MeV \cite{18},
energy loss measurements of different ions \cite{19}, and
the $^4$He($^7$Li,$^4$He)$^7$Li$^*$
reaction at ${\rm E_{lab}}= 3.325$~MeV  \cite{20}.
The weighted average of
the results yields N$_{^4{\rm He}}=4.21\pm0.14\times10^{17}$~atoms/cm$^2$
with an effective target length of $42.6\pm1.4$~mm, that corresponds
to an energy loss, E$_{\rm loss}$, for the $^{12}$C ions smaller than 
25 keV in the investigated energy range. The small target thickness
leads to a nearly constant cross 
section along the target, thus E$_{\rm eff}$ = (E$_{\rm beam}-{\rm E_{loss}}$/2)*M$_{^4{\rm He}}$/(M$_{^{12}\rm C}+\rm M_{^4{\rm
He}}$).

\begin{figure*}
 \vspace{-1cm}
  \center\includegraphics[angle=0,width=13cm]{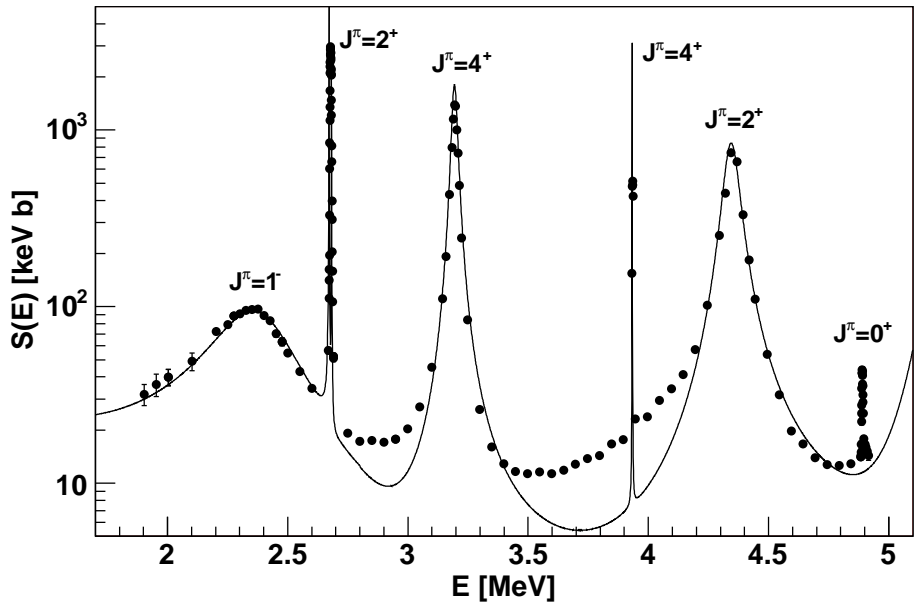}
  \caption{\label{sfactor}Total S(E) factor of the reaction $^{12}\rm C(\rm \alpha,\gamma)^{16}\rm O$.
  Data near the narrow resonances at $\rm E=2.68$, 3.9, and 4.9~MeV
  are thick-target yields.
  The solid line represents the sum of the different amplitudes extracted from 
  a recent R-matrix calculation \cite{k1}. Error bars shown are statistical only.}
\end{figure*}

The number of projectiles N$_{^{12}\rm C}$ is determined through
the detection of the elastically scattered $^4$He nuclei in two 
collimated silicon detectors located in the target chamber at
75$^\circ$ from the beam axis. Calibration runs performed at 
each energy before and after the measurement runs allowed
to relate the observed scattering rate to the concurrent beam current.
The scattering rate was measured in short runs of typically 60 s to achieve a
statistical precision of better than 1\%. The beam
current without target gas was monitored in a Faraday cup (FC2) located after
the quadrupole triplet before and after the determination of the
scattering rate. This procedure was found to be independent of the beam focussing
and reproducible within the statistical error. A 100\% transmission of
the incident beam through the
gas target is a requirement of the separator \cite{18} and was verified by
the full transmission through a retractable focussing aperture in front
of the gas cell with a diameter of 3~mm.

\begin{table}
\caption{\label{table1}Astrophysical S(E) factor of
$^{12}\rm C(\rm \alpha,\gamma)^{16}\rm O$. The quoted errors are statistical
only. The data in the energy range of the narrow resonances are left out.}
\begin{tabular}{cccc}
\hline
E [MeV] & S [keVb] & E [MeV] & S [keVb] \\
\hline
1.903	& $31.8	\pm4.3	$ &	3.249	& $84.2	\pm2.5	$ \\
1.953	& $36.3	\pm5.3	$ &	3.298	& $26.1	\pm0.9	$ \\
2.003	& $39.9	\pm4.3	$ &	3.348	& $16.1	\pm0.5	$ \\
2.102	& $49.0	\pm5.6	$ &	3.398	& $12.9	\pm0.5	$ \\
2.202	& $72.3	\pm2.7	$ &	3.448	& $11.7	\pm0.4	$ \\
2.252	& $79.1	\pm2.9	$ &	3.498	& $11.3	\pm0.4	$ \\
2.277	& $88.4	\pm3.0	$ &	3.548	& $11.6	\pm0.4	$ \\
2.302	& $90.9	\pm3.3	$ &	3.597	& $11.3	\pm0.4	$ \\
2.327	& $95.1	\pm3.0	$ &	3.647	& $11.9	\pm0.3	$ \\
2.352	& $96.4	\pm3.0	$ &	3.697	& $12.8	\pm0.4	$ \\
2.376	& $96.7	\pm2.9	$ &	3.747	& $13.8	\pm0.4	$ \\
2.402	& $88.8	\pm2.5	$ &	3.797	& $14.3	\pm0.4	$ \\
2.426	& $83.0	\pm2.7	$ &	3.847	& $16.6	\pm0.5	$ \\
2.451	& $70.2	\pm2.9	$ &	3.896	& $17.6	\pm0.5	$ \\
2.476	& $63.3	\pm3.1	$ &	3.946	& $23.1	\pm0.7	$ \\
2.476	& $63.3	\pm3.1	$ &	3.996	& $23.7	\pm0.7	$ \\
2.501	& $54.7	\pm1.7	$ &	4.046	& $29.5	\pm0.9	$ \\
2.551	& $42.8	\pm1.5	$ &	4.096	& $34.2	\pm1.0	$ \\
2.601	& $34.4	\pm1.1	$ &	4.145	& $41.3	\pm1.1	$ \\
2.750	& $19.2	\pm0.7	$ &	4.195	& $57.0	\pm1.7	$ \\
2.800	& $17.3	\pm0.7	$ &	4.245	& $101.7\pm2.1	$ \\
2.850	& $17.4	\pm0.6	$ &	4.295	& $253.8\pm4.1	$ \\
2.900	& $17.1	\pm0.5	$ &	4.320	& $440	\pm11	$ \\
2.949	& $17.7	\pm0.6	$ &	4.345	& $744	\pm11	$ \\
2.999	& $20.3	\pm0.6	$ &	4.370	& $661	\pm13	$ \\
3.049	& $27.0	\pm0.9	$ &	4.395	& $331.5\pm5.5	$ \\
3.099	& $45.3	\pm1.4	$ &	4.420	& $183.4\pm4.9	$ \\
3.144	& $110.6\pm3.3	$ &	4.445	& $110.5\pm2.5	$ \\
3.158	& $191.9\pm6.0	$ &	4.494	& $53.6	\pm1.6	$ \\
3.174	& $432	\pm11	$ &	4.545	& $31.6	\pm0.9	$ \\
3.184	& $797	\pm14	$ &	4.594	& $19.7	\pm0.6	$ \\
3.189	& $1156	\pm21	$ &	4.644	& $16.7	\pm0.5	$ \\
3.194	& $1392	\pm21	$ &	4.694	& $14.0	\pm0.4	$ \\
3.199	& $1363	\pm13	$ &	4.744	& $12.7	\pm0.4	$ \\
3.204	& $1001	\pm17	$ &	4.793	& $12.6	\pm0.4	$ \\
3.209	& $740	\pm17	$ &	4.843	& $12.9	\pm0.4	$ \\
3.214	& $484.6\pm8.2	$ &	4.883	& $14.1	\pm0.4	$ \\
3.224	& $244.5\pm5.8	$ &	4.917	& $14.3	\pm0.8	$ \\
\hline
\end{tabular}
\end{table}

The charge state distribution $\Phi_{\rm R}$ of the $^{16}\rm O$
recoils produced in the $^4$He gas target depends on the geometric origin in 
the target: $^{16}\rm O$ recoils produced in the upstream part
of the target will most likely reach an equilibrium charge state 
distribution in the passage of the remaining target length, while
those produced near the downstream end of the 
target will not, i.e. they will keep memory of the charge
state at the moment of their formation. Since this effect
is not accurately predictable \cite{19}, the charge state
distribution of the $^{16}\rm O$ recoils after passing the He gas is affected by a significant 
uncertainty. To remove this uncertainty, an Ar 
stripper was installed after the $^4$He gas target
with a number density ${\rm n_{Ar}} = 5.6\pm0.6\times10^{16}$~atoms/cm$^2$.
This density is sufficient for all ions produced at different locations in
the $^4$He target to reach the same charge state distribution.
Finally, the observed $^{16}\rm O$ charge state distribution in the combination
of $^4$He and Ar gas was measured over the full energy
range \cite{20}: for each charge
state the separator was set properly and the resulting current observed at the
end of the separator in FC4 (fig. \ref{setup}). The measured charge state
distribution differs from the equilibrium charge state distribution of
$^{16}$O ions in Ar gas \cite{19} due to the effect of charge exchange in the $^4$He
rest gas in the downstream pumping stages of the gas target after the
post-target stripper. This effect
is energy and charge state dependent and amounts to a 
chance of the equilibrium charge state distribution of 20~\% at most.

The transmission of the separator T$_{\rm RMS}$ essentially
depends on its acceptance compared to the emittance of the 
recoils, which in turn depends on $\gamma$-ray emission, target
effects, and the beam emittance. The angular acceptance of 
ERNA has been measured using an $^{16}\rm O$ beam and an
electrostatic deflection unit \cite{18}, which can deflect the beam at 
any position within the target region in order to simulate
the recoil ion angular opening at the different geometrical 
locations where $^{16}\rm O$ recoils are produced.
The energy acceptance was measured by varying the beam 
energy from the accelerator. For both quantities, ERNA
turned out to fulfill the requirements for measuring 
$\sigma({\rm E_{eff}})$  in the energy
range $\rm E=1.3$ to 5.0~MeV within a region of
$\pm70$~mm around the target center, that 
includes 96\% of the target nuclei. The angular straggling 
of the recoils in the gas target results in a loss of recoils
less than 0.5~\%, which gives a final value of ${\rm T_{RMS}}=1_{-0.01}^{+0.0}$ \cite{20}.
The detector efficiency $\epsilon_{\rm d}=1$ is reduced by the 
transparency T$_{\rm PGAC}$ of the grids of a parallel grid avalanche counter
at the entrance window of the $\Delta$E-E telescope
leading to a total detection efficiency $\epsilon=\epsilon_{\rm d}{\rm
T_{PGAC}}=0.980\pm0.005$.

It is important to note that the 100~\% transmission of
the recoils in the selected charge state is a key requirement for 
measuring reliably the absolute capture cross section $\sigma({\rm E_{eff}})$
using a recoil mass separator. Indeed, this is a condition for 
determining $\sigma({\rm E_{eff}})$ without a detailed knowledge of the distribution
of the recoils in the phase space, which depends 
on the $\gamma$-ray energy and $\gamma$-ray angular distribution, on
target effects and beam emittance. 

Finally, background runs (fig. \ref{matrix} lower panel) were performed at
each energy using the Ar post-target stripper only. The 
measurements showed the presence of an $^{16}\rm O$ background,
where at energies above 2.0~MeV a signal-to-background 
ratio of typically 20 was observed.
The observed background rate in the region of interest 
was normalized to the number of projectiles and subtracted at each energy. 
An investigation
of this $^{16}\rm O$ background excluded as possible sources a 
beam contamination, which is suppressed by the ERNA beam
purification system, as well as elastic scattering on 
rest gas. There are strong indications that the
background is mostly due to the fraction of the $^{12}$C beam 
impinging on the plates of the first Wien filter, and possibly
other beamline components, producing $^{16}\rm O$ via 
$^{12}$C+$^{12}$C fusion, e.g. the $^{12}$C($^{12}$C,$^8$Be)$^{16}$O
reaction. Since the fraction of the beam
impinging on the plates increases with decreasing energy, 
measurements at energies lower than $\rm E =2.0$~MeV are
affected by a signal-to-background ratio significantly 
poorer than at higher energy.  

\section{Results and conclusions}

Fig. \ref{sfactor} and table \ref{table1} show the total S(E) factor at
$\rm E =1.9$ to 4.9~MeV obtained with ERNA; the statistical error
is of the order of 4\% or less at energies above 2.2~MeV and the systematic
error is 6.5\% (3\% N$_{^4{He}}$, 2\% N$_{^{12}C}$, 1\% T$_{RMS}$, 2\% $\Phi_R$, 0.5\%
$\epsilon$). 
In fig. \ref{sfactor} we present also a comparison of our data with the sum
of the different amplitudes as reported in \cite{k1}, i.e. the ground state
transition (E1 and E2) plus cascade transitions.
Note the good agreement in absolute S(E) values between previous work
and the present data on top of the $\rm E=2.4$, 3.2, and 4.3~MeV resonances. 
The R-matrix calculation represents the best fit to
the available experimental data at that time including resonance information from \cite{24}. One should note
that in this analysis the interference effects of the cascade transitions
are neglected. There is a clear
disagreement at energies around $\rm E=3.0$ and 4.0~MeV, where the
calculation of \cite{k1} underestimates the total cross 
section, while it slightly overestimates the cross
section on the high energy tail of the broad
J$^\pi$=1$^-$ resonance at $\rm E=2.4$~MeV. 

A preliminary analysis indicates that these discrepancies may be partly caused by a wrong choice of
the interference pattern of the different amplitudes, especially of the E2 amplitude. The analyses
\cite{k1,k2} show that the $\chi^2$-fit is hardly sensitive to different E2 interference patterns
leading to a broad range of extrapolated ${\rm S_{E2}(E_0)}$ values. 
A detailed study of the influence of the present data on S(E$_0$)
is in progress, that requires an R-matrix code including a fit to 
the different amplitudes (see for example \cite{22,k1}) as well as to the total cross section.

Finally, a new resonance was found in $^{12}\rm C(\rm \alpha,\gamma)^{16}\rm O$
at $\rm E=4.888$~MeV, corresponding to a known 0$^+$ state
($\Gamma=1.5\pm0.5$~keV) reported in \cite{24}. The analysis of the
recoil data results in a resonance strength $\omega\gamma=11.2\pm1.5$~meV.

In conclusion, the ERNA approach provided new data of
$^{12}\rm C(\rm \alpha,\gamma)^{16}\rm O$ at energies above 1.9~MeV,
which are needed for an improved S(E$_0$) determination.

The authors thank the technical staff of the Mechanical Workshop and of the Dynamitron Tandem Laboratorium in 
Bochum for their support to the project.
The project is supported in part by Deutsche
Forschungsgemeinschaft (Ro429/35-2) and Istituto
Nazionale di Fisica Nucleare (ERNA), and DAAD-CRUI VIGONI.

\end{document}